\documentclass[12pt]{article}
\usepackage{amssymb}
\usepackage{euscript}

\usepackage[latin1]{inputenc}
\usepackage[T1]{fontenc}
\usepackage{amsmath}
\usepackage{amsfonts}
\usepackage{amssymb}
\usepackage{latexsym}
\usepackage[english]{babel}
\usepackage[dvips]{graphicx,color}
\usepackage{latexsym}
\usepackage{graphicx}
\usepackage{graphics}
\usepackage{pstricks}

\begin{document}

\newcommand{\be}{\begin{equation}}
\newcommand{\ee}{\end{equation}}
\newcommand{\epe}{\end{equation}}
\newcommand{\bea}{\begin{eqnarray}}
\newcommand{\eea}{\end{eqnarray}}
\newcommand{\ba}{\begin{eqnarray*}}
\newcommand{\ea}{\end{eqnarray*}}
\newcommand{\epa}{\end{eqnarray*}}
\newcommand{\ar}{\rightarrow}

\def\A{\tilde{a}}
\def\G{\Gamma}
\def\r{\rho}
\def\D{\Delta}
\def\R{I\!\!R}
\def\l{\lambda}
\def\D{\Delta}
\def\d{\delta}
\def\T{\tilde{T}}
\def\k{\kappa}
\def\t{\tau}
\def\f{\phi}
\def\p{\psi}
\def\z{\zeta}
\def\ep{\epsilon}
\def\hx{\widehat{\xi}}
\def\a{\alpha}
\def\b{\beta}
\def\O{\Omega}
\def\M{\cal M}
\def\g{\hat g}
\newcommand{\dslash}{\partial\!\!\!/}
\newcommand{\aslash}{a\!\!\!/}
\newcommand{\eslash}{e\!\!\!/}
\newcommand{\bslash}{b\!\!\!/}
\newcommand{\vslash}{v\!\!\!/}
\newcommand{\rslash}{r\!\!\!/}
\newcommand{\cslash}{c\!\!\!/}
\newcommand{\fslash}{f\!\!\!/}
\newcommand{\Dslash}{D\!\!\!\!/}
\newcommand{\Aslash}{{\cal A}\!\!\!\!/}

\hspace{11cm} CERN-PH-TH/2011-235
\vspace{3mm}

\begin{center}

\vspace{3mm}



{\large Emergent spacetime, and a model for unitary gravitational collapse in AdS}

\vspace{.3in}
Marcelo Botta Cantcheff $^{\dag\,\ddag}$\footnote{e-mail:
bottac@cern.ch, botta@fisica.unlp.edu.ar}

\vspace{.2 in}

$^{\dag}${\it IFLP-CONICET CC 67, 1900,  La Plata, Buenos Aires, Argentina}\\

$^{\ddag}${\it CERN, Theory Division, 1211 Geneva 23, Switzerland}
\vspace{.4in}

\end{center}

\begin{abstract}
\noindent

We propose a CFT unitary description of the gravitational collapse. The starting point is the model of a black hole in AdS proposed by Maldacena in Ref. \cite{eternal}. We show that by proposing a two-copies version of the AdS/CFT conjecture, the process of formation of black holes so as other spacetimes with horizons may be described as an unitary process in the dual field theory. In doing this, we construct a well defined framework to describe general
 spacetimes as entangled states, in terms of the spectrum of states
   on the exact Anti-de-Sitter background. As application, we show how the description of the Hawking-Page transition results simplified in this formalism and some novel aspects may be observed.
Finally, a simplified analysis based on weakly coupled bulk fields is discussed.

~


\end{abstract}

\section{Introduction}

The AdS/CFT correspondence represents the paradigmatic case where spacetime may be defined
 as emergent from a gauge theory on its conformal boundary \cite{adscft}.
 An exact description of quantum gravity is expected to be constructed in general based
 on this and other examples of holography, however, we do not have yet a deep description
  of the how the states of the spacetime and gravitational dynamics emerge from the
   degrees of freedom of quantum field theory \cite{llm}\cite{seiberg}.
 From another perspective, although the unitary evolution of gravitational systems is a foundational property of this holographic approach, the problem of describing the gravitational collapse into black holes has not been clarified so far \cite{infoBH}, even in conventional AdS/CFT \cite{bala-info}.

 Consider the simplest scenario of the correspondence: a $M_0 \equiv AdS_{5} \times S^5$ spacetime, whose boundary is $B \sim S^3 \times {\mathbb R}$. It states that a complete quantum gravity (and quantum field theory)\footnote{Probably formulated as a string theory} on this background $M_0$ is equivalently described by CFT (with a hamiltonian $H$) on the boundary $B$; and the corresponding quantum states belong to the Hilbert space ${\cal H}_{CFT}$. The standard interpretation is that the bulk geometry $M_0 \sim AdS_{5} \times S^5$, corresponds to the fundamental state $\left|0\right\rangle$ in this space.
The main issue of the holographic formulation of quantum gravity (or, the so-called spacetime emergence), is how to describe more general background geometries with the same asymptotic behavior; it is often assumed however that general spacetime backgrounds should be described by non trivial choices of that fundamental state. We shall revisit this interpretation by considering a more general structure.

On the other hand, Maldacena emphasized that a non trivial, asymptotically AdS geometry as an eternal black hole in an AdS box, shall be described by an (entangled) state \cite{eternal}:
\be \left.\left|0(\beta)\right\rangle \! \right\rangle = \sum_n \,\frac{e^{-\b E_n}}{Z^{1/2}} \left|n\right\rangle \otimes \left|\tilde{n}\right\rangle
~~\b\equiv(k_B T)^{-1}\label{BHstate}, \ee in $ {\cal H}\otimes\widetilde{{\cal H}}$, where ${\cal H}$ is the Hilbert space of $CFT$ and $\widetilde{{\cal H}}$ an identic copy; $E_n$ are the eigenvalues of the CFT Hamiltonian and is $\left|n\right\rangle$ a complete basis of eigenstates ($T$ is the Hawking temperature of the black hole). In a recent essay \cite{VR}, it was properly appreciated that the holographic emergence of the spacetime should be intimately related to quantum entanglement.
We are going to propose a model of unitary evolution into this state by conciliating these facts.

 In this work we will argue that the process of formation of AdS-Black Holes may be described by an unitary evolution operator, by assuming that
 any spacetime with certain asymptotics, is given by a general vacuum state of decoupled quantum systems (so as (\ref{BHstate})), in agreement with the arguments of \cite{eternal} \cite{VR} \cite{israel}.

The paper is organized as follows: in Section 2, we extend the CFT theory/operators according to the canonical rules of Thermofield Dynamics \cite{tu}\cite{ume} , in order to define the more general asymptotically AdS states of the spacetime, with two disconnected boundaries. This is done in Section 3 through formal statements which extend the conventional AdS/CFT duality. In Section 4 we argue general unitarity of the gravitational collapse,
describe the process, and discuss a canonical example; furthermore, we observe that the thermodynamic laws determine the final state, and consequently, it is proposed that the gravitational forces may be holographically related to thermodynamics. In Section 5 we give a simple description of the Hawking-Page transition in this new language. Finally, a simplified approach based on canonically quantized (bulk) fields, is discussed in Section 6. Concluding remarks are collected in Section 7.

\section{Duplication of CFT from TFD rules}

Expression (\ref{BHstate}) is a thermal state in the TFD formalism \cite{tu}\cite{ume}.
According to it one shall consider \emph{a copy} of the CFT theory invoked by the Maldacena conjecture described by a Hamiltonian $\tilde{H}$,
 and defined on an identical copy of the manifold $B$, denoted by $\tilde{B}$ (the conformal boundary of a identical copy of the bulk $\widetilde{M}$).
 In conventional TFD this copy is a fictitious (non-physical) system, however according to the arguments put forth by Israel \cite{israel}, it may be interpreted as a real system causally disconnected from the first one. There are precise rules to construct such a duplicated theory, the "tilde
conjugation rules", or simply TFD rules \cite{kha5}:
\begin{eqnarray}
( X Y)\widetilde{}
&=&\widetilde{X}\widetilde{Y}, \nonumber \\
(c\, X + Y)\widetilde{} &=&c^{\ast
}\,\widetilde{X}_{i}+\widetilde{Y}_{j}, \nonumber \\
(X^{\dagger })\widetilde{}
&=&(\widetilde{X})^{\dagger }, \nonumber \\
(\widetilde{X})\widetilde{} &=& X,  \nonumber \\
\lbrack \widetilde{X},Y] &=&0 \, ,  \label{til}
\end{eqnarray}
for arbitrary operators of CFT. This structure is related to a $c^\star$-algebra, and the rules (\ref{til}) may be identified with the modular conjugation of the standard representation \cite{emch}.

 So, we can define the conformal theory $CFT^2$ as the direct product of both (schematically $CFT \otimes \widetilde{CFT}$), whose time evolution is given by a decoupled theory $\hat{H} \equiv H-\tilde{H}$ and
the total Hilbert space is the tensor product of the
two state spaces $ {\cal H}_{CFT^2} \equiv {\cal H}_{CFT}\otimes {\cal H}_{\widetilde{CFT}}$.

The states in this doubled space include all the \emph{statistical} information of the system CFT on behalf of
  density operators. According to several remarks \cite{macrogeom}, which enforce the motivations of the present proposal, this information is essential to get a complete description of the spacetime geometry.

\section{The AdS/CFT$^2$ correspondence}

In what follows, we are going to focus on general spacetimes geometries whose asymptotic boundaries have the local structure $AdS_{d+1} \times {\cal N}$, where ${\cal N}$ denotes a compact Riemannian manifold.
For simplicity, we often refer to the well known standard $AdS_{5} \times S^5$ asymptotics.

The main hypothesis which our construction is based on, may be expressed as follows.


\vspace{0.7cm}


\textbf{Assumption I.1} A general spacetime $M$ with asymptotic behavior $AdS_{5} \times S^5$ near two conformal boundaries, is described by an state
\be \left.\left|M\right\rangle \! \right\rangle = \sum_{n, \tilde{m}} \, G_{n\tilde{m}} \left|n\right\rangle \otimes \left|\tilde{m}\right\rangle\,,
\label{geometrystate} \ee
where $G_{n\tilde{m}}$ are complex numbers, and $\left|n\right\rangle \otimes \left|\tilde{m}\right\rangle\,$ is a complete basis of $ {\cal H}_{CFT^2} = {\cal H}_{CFT}\otimes {\cal H}_{\widetilde{CFT}}$.

\vspace{0.7cm}


\textbf{Assumption I.2} A conformal field theory $CFT^2 \sim CFT\otimes \widetilde{CFT}$ (as described above), in the ground state (\ref{geometrystate}); is \emph{equivalent} (dual)
to a complete quantum field theory (including gravity) on a spacetime $M$ with two conformal boundaries, and $AdS_{5} \times S^5$ asymptotics.

\vspace{0.7cm}

This statement have meaningful implications on the mechanism of emergence of the spacetime and gravity. In particular \textbf{I.1} can be viewed as a definition of the \emph{quantum states of the geometry} with fixed asymptotics. In principle this rule may be generalized to other asymptotic behaviors (and dual field theories), than the standard $AdS_{5} \times S^5$-case studied here.

As pointed out above, the state (\ref{geometrystate})
 encodes the quantum and \emph{statistical} information about the CFT system \cite{macrogeom}.
Recently, an illuminating work observed the necessary relation between the spacetime geometry and the entanglement in the context of holography \cite{VR}.
Our proposal is in agreement with these arguments since the general states of the geometry are regarded, precisely, as entangled states.
Here we give a constructive step by proposing a \emph{rule} to describe general spacetimes (with two AdS asymptotic regions) in the context of holography.

Notice that the general claim above describes the more general spacetimes with \emph{two} asymptotic boundaries. The conventional case with only one, is implicitly included in this definition as two disconnected copies of globally AdS-spacetime, or their respective (solitonic) deformations \cite{llm}.
Separable states are in fact expressed as
\be \left.\left|M\right\rangle \! \right\rangle =  \left|M_1\right\rangle \otimes |\,\widetilde{M}_{2}\,\rangle\,
\label{separablestate} .\ee
Then the correlation functions of arbitrary operators $O_1 , \tilde{O}_2$, in $CFT$ and $\widetilde{CFT}$ respectively, are
\be \left\langle \!\left\langle M \right|\right. O_1 \, \tilde{O}_2\left.\left|M\right\rangle \! \right\rangle = 0
\label{separablecorrel} .\ee
Therefore, as argued in \cite{VR} this state $\left.\left|M\right\rangle \! \right\rangle$ describes a \emph{disconnected geometry} given by the union of \emph{two} asymptotically AdS spacetimes with one conformal boundary\footnote{Since there is no finite length curves connecting the boundaries of $M_1$ and $\tilde{M}_2$.}. So topological connectivity shall be a property of entangled states in the $ {\cal H}_{CFT^2} $ space.

In particular, those geometry states whose conformal boundaries are connected by an Einstein spacetime, necessarily contains one horizon separating them \cite{galloway}. Then, an interesting question that arises is if it may be generalized/extended to multiple $n$ disconnected boundaries (not only the obvious extension to multiple pairs ($n\sim 2m$) of $CFT^2\sim CFT\otimes \widetilde{CFT}$ by simply taking $m$ copies of the present structure), and if it would actually be describing something else.
If yes, it shall presumably be a description of general spacetimes with $m$ disconnected horizons.

\subsection{The spacetime picture}

One of the main aspects of Assumption \textbf{I} is what it implies on the form and structure of the quantum states of the geometry. If we assume that the Type IIB string theory on global AdS is solvable, and the corresponding Hilbert space ${\cal H}_{1 String}$, whose basis elements schematically read $\left| \{N_n\}_n \right\rangle_{} = \bigotimes_{n=-\infty}^\infty(A_{n}^{\dagger\:\mu})^{N_n} \left|0\right\rangle$, is well defined\footnote{Then one may formally define the string Fock space as ${\cal F}_S\equiv\left|0\right\rangle \bigoplus_{m=1}^\infty {\cal H}_{m String} $}. One can formulate this theory as a set of fields of embedding of the closed string into the AdS spacetime, and then to expand all fields in Fourier modes on the circle labeled by the parameter $\sigma$; for each Fourier mode
we get a harmonic oscillator. So formally, $A_{n}^{\dagger\:\mu}$ denotes the corresponding creating operators of string excitations upon quantization; $n\in \mathbb{Z}$ is the label of the Fourier mode $N_n$
denotes the total occupation number of that mode, including bosons and fermions (here we consider only the bosonic sector for simplicity); the string vacuum is naturally identified with the pure spacetime state $\left|0\right\rangle\equiv\left|M_0\right\rangle$.
 For instance, the spectrum is exactly solvable in the pp wave limit (Ref. \cite{pp}), where the AdS/CFT dictionary is known \cite{BMN}.
In this sense, AdS/CFT implies the existence of an invertible map:
$\left|n \right\rangle_{CFT} \mapsto \bigotimes_{n}(A_{n}^{\dagger\:\mu})^{N_n} \left|M_0\right\rangle$;
using it, we then may write down \emph{any} state of the spacetime (with the same asymptotics as the global AdS) as a quantum superposition of these states:
\be \left.\left|M\right\rangle \! \right\rangle = \sum_{\{N _n\}\{\tilde{N}_m\}} \, \mu^{(bulk)}_{\{N _n\}\{\tilde{N}_m\}} \, \left| \{N_n\}_n \right\rangle_{}\,\otimes\, | \{\, \widetilde{N}_m \,\}_m \rangle_{} \,~,~~~\mu^{(bulk)}_{\{N _n\}\{\tilde{N}_m\}}\in \mathbb{C}
\label{gstate-G} .\ee
In addition, one can expect that this state is the ground state of the Fock space of strings quantized on the corresponding geometry $M$ (whenever it makes sense), and the coefficients $\mu$ should encode the Bogoliubov map between this and ${\cal H}_{1 String}$.
The useful conclusion is that, in principle, we may have \emph{a class} of spacetimes (with AdS asymptotics), and their corresponding quantum excitations, just by knowing as to quantize the string in the simplest exact AdS geometry. This paradigm provides the detailed structure of the states of the geometry, and shall be generalizable to other $M_0$-asymptotics \footnote{A related approach was constructed in Ref. \cite{closedpure} to describe Dp-branes as vacuum states }. A stimulating consequence is the possibility of describing different non trivial topologies in this way.

As pointed out before, disentangled states as
\be \left.\left|M\right\rangle \! \right\rangle = \sum_{\{N_n\}\{\tilde{N}_m\}} \, \mu^{}_{\{N _n\}}\, \tilde{\mu}_{\{\tilde{N}_m\}} \, \left| \{N_n\}_n \right\rangle_{}\otimes |  \{\widetilde{N}_m \}_m \rangle_{} \,=\left|M_1\right\rangle \otimes |\,\widetilde{M}_{2}\,\rangle\,
\label{gstate-G-sep} ,\ee
describe a \emph{disconnected geometry} given by the union of two spacetimes, asymptotically AdS.
If our conclusion above is correct, stationary states (\ref{gstate-G-sep}) shall correspond to solitonic classical solutions of (super-)gravity.  In a forthcoming work we will study how classical geometries are recovered in this framework \cite{forth}, and furthermore, how the standard holographic GKPW map \cite{GKPW} is reformulated.

\section{Collapse}

Notice that this generalizing formulation properly contains the eternal Black Hole description explained above as a particular case: for the Black Hole state, the coefficients $G_{n\tilde{m}}$ are given in (\ref{BHstate}). The model that we are proposing here is based on the known fact that in a canonical quantum field theory, a pure (disentangled) state as $\left.\left|0\right\rangle \! \right\rangle \equiv \left|0\right\rangle \otimes \left|\tilde{0}\right\rangle $ is connected to (\ref{BHstate}) by an unitary operator in a finite volume system \cite{ume}. Therefore, our main claim about the unitary collapse is proven in the following sense:

~

\textbf{(i)} A Black Hole state (\ref{BHstate}) may be achieved from a pure AdS-state $\left.\left|0\right\rangle \! \right\rangle$ by an unitary transformation.

\textbf{(ii)} This is referred to as a Bogoliubov transformation, which in particular, constitutes the simplest example of \emph{evolution operator} to describe the collapse process\footnote{It will be shown in an example below. The reader should interpret this remark as addressed to emphasize the \emph{existence} of an effective model for the unitary collapse dynamics, rather as a description of the real process.}.

~

In the standard version of AdS/CFT, the claim \textbf{(i)} is automatically true by virtue of Assumption \textbf{I}, because of the simple structure $S^{d-1} \times {\mathbb R}$  of the boundary manifold where the dual CFT theory is defined. So in principle, other asymptotic conditions of the spacetime can be put into this scheme, provided that the dual theory is a well defined QFT whose degrees of freedom lie on a space of finite volume.

The validity of these statements is general, and does not depend on the specific form of the field theory or its interacting structure (however, a canonical realization of this scenario is discussed below). In this general context, we can do a further observation that signals the generality and ``unavoidability'' of the collapse, and shed light on the very interpretation of the process in terms of quantum statistics:

~

\textbf{(iii )} In the TFD formalism, the thermodynamic entropy (or the free energy in the canonical ensemble) is an extremum in the final state (\ref{BHstate})
\cite{tu,ume}, so it may be claimed that \emph{the physical cause underlying the gravitational collapse, is the tendency of the system to achieve equilibrium states}.

~

This agrees indeed with the interpretation of gravity as an entropic force \cite{verlinde, macrogeom}. But even thought that this is a general feature, an example of the microscopic description of a dynamical process that ends in the Black Hole state, does exist; and it may be described as a Bogoliubov propagator in the sense \textbf{(ii)}.

These observations may be collected in a supplementary assumption related to the \emph{evolution} of holographic spacetimes, which essentially claims that the gravitational collapse so as general gravitational processes, are holographically ruled out by the thermodynamics laws:

~

\textbf{Assumption II.1:} Thermodynamics of the $CFT$ theory is exactly described by the TFD formalism \cite{tu,ume}. In particular, there exist an entropy operator denoted by $K$ \cite{tu}, defined such that its expectation value in the state (\ref{geometrystate}) measures the entropy of entanglement between the system and its conjugate sector $\widetilde{CFT}$, and coincides with the \emph{thermodynamic} entropy $S$ of the system at thermal equilibrium (divided by the Boltzmann's
constant \cite{tu}).

\textbf{Assumption II.2:}
The stationary states of the geometry correspond to equilibrium configurations; i.e. the quantity:
\be
S= \frac{1}{k_{B}}\left\langle\! \left\langle M(\theta) \right|\right.K \left.\left|M(\theta)\right\rangle\! \right\rangle
\ee
is an extreme (maximum) under arbitrary variations of proper parameters $\theta$.

~

Consistency of these statements with the Ryu-Takayanagi holographic formula \cite{takaya} requires that
the geometry $M(\theta)$ contains an extremal surface $\mathcal{S}$ ending on the conformal boundary, whose area is $S$ in the limit as the surface $\partial\mathcal{S}$ encloses all the degrees of freedom of dual boundary theory (CFT).

The paradigmatic example for this type of picture is an oscillator-like system, where the generator algebra of the Bogoliubov transformations is known.
In Section 6 we argue that this may be related to a perturbative description of the bulk fields, quantized on a global AdS background geometry, and whose Fock space is canonically built with the normalizable modes \cite{giddings-gary}.
 In such a sense, we may interpret what follows as a quantitative (bulk) description of this scenario, in the proper context.

 In these systems
the Hamiltonian operator is
\be\label{hamcan}
H= : \sum_{n,I} w_{n,I} \, a_{n}^{\:I}a_{n}^{\dagger\:I} \,:\,,
\ee
where $n$ denotes the set of numbers characterizing the discrete (positive) frequency modes, and the indices $I,J,...=1,....,f $ label the independent physical fields.
$a_n^I, a_n^{\dagger \:I} $ are conventional creation/annihilation operators, which must extended according to the TFD rules as prescribed by \textbf{I.2}. They satisfy the extended algebra:
\begin{eqnarray}
\left[a_{n}^I,a_{m}^{\dagger \:J}\right]
&=&\left[\tilde{a}^{I}_{n},\tilde{a}_{m}^{\dagger\:J}\right]
=  \delta_{n,m}\delta^{I,J},\label{alg}
\nonumber
\\
\left[a_{n}^{\dagger\:I},\tilde{a}_{m}^{J}\right]
&=&\left[a_{n}^{\dagger\:I},\tilde{a}_{m}^{\dagger\:J}\right]
=\left[a_{n}^{I},\tilde{a}_{m}^{J}\right]=
\left[a_{n}^{I},\tilde{a}_{m}^{\dagger\:J}\right]=0.
\end{eqnarray}
 The ground state in this extended theory is conventionally defined by
\be {a}^{I}_{n}\left.\left|0\right\rangle \! \right\rangle= \A_{n}^I
\left.\left|0\right\rangle \! \right\rangle = 0\,\,\,\,. \ee
Let us consider Bogoliubov transformations of this extended theory, such that:
\bea a_{n}^{I}(\theta_{n}) &=& e^{-iG}a_{n}^{I}e^{iG}
=\cosh(\theta_{n})a_{n}^{I} - \sinh(\theta_{n}){\widetilde
a}_{n}^{\dagger \: I}
\\
\widetilde{a}_{n}^{ I}(\theta_{n}) &=& e^{-iG}{\widetilde
a}_{n}^{ I}e^{iG}
= \cosh(\theta_{n}) a_{n}^{I} -  \sinh(\theta_{n}) {\widetilde
a}_{n}^{\dagger \: I}
\eea
As the Bogoliubov transformation is canonical, the transformed
operators obey the same commutation algebra (\ref{alg}).
 These operators annihilate the state
written in $(\ref{tva})$ defining it as the vacuum
\begin{eqnarray}
a_{n}^{I}(\theta_{n})\left |0(\theta)\right\rangle &=& \widetilde
{a}_{n}^{I}(\theta_{n})\left |0(\theta)\right\rangle = 0,
\end{eqnarray}
\be \left.\left|0(\theta)\right\rangle\! \right\rangle =
e^{-i{G}} \left.\left|0\right\rangle \! \right\rangle  \label{tva}. \ee
Then, the physical Fock space is constructed by applying the
 creation operators $
a_{n}^{\dagger \: I}(\theta_{n}), {\widetilde
a}_{n}^{\dagger \: I}(\theta_{n})$ to the vacuum $(\ref{tva})$.
 According to the arguments above, in the present framework we shall identify these states with the spacetime geometries $M(\theta)$ itself.

 Requiring an extra symmetry for the state $\widetilde{\left.\left|0(\theta)\right\rangle\! \right\rangle}= \left.\left|0(\theta)\right\rangle\! \right\rangle $, and that $e^{-i{G}}$ be unitary,
the generator for these transformations is Hermitian and given by: \be G= -i
\,\,\sum_{n, I}\theta_{n}\,(a_{n}^{I} \tilde{
a}_{n}^{I}\, - \,\tilde{a}_{n}^{\dagger\:I} a_{n}^{\dagger\:I} )\,.
\ee By using a normal ordering, we may verify:
 \be \left.\left|0(\b)\right\rangle\! \right\rangle =
e^{-i{G}} \left.\left|0\right\rangle \! \right\rangle =
\prod_{n, I}\left[\left( \frac{1}{\cosh(\theta_{n})}\right)^f
e^{\tanh(\theta_{n})\,\,\,a_{n}^{\dagger\:I} {\tilde
a}_{n}^{\dagger\:I}} \right]
\left.\left|0\right\rangle\!\right\rangle \label{tva2}. \ee
where $Z^{-1/2}= \prod_{n}\left( \frac{1}{\cosh(\theta_{n})}\right)^f$, is a well defined real number for a finite volume system.
The precise value of the parameters in a AdS Black Hole is:
 \be\label{distr-equil}\tanh(\theta_{n}) = e^{-n\b/2},\ee
that corresponds to the values that maximize the thermodynamic entropy with respect to the parameters $\theta$`s. In the TFD formalism, the thermodynamic entropy is canonically defined as the expectation value of the operator:
\begin{eqnarray}
K &=& - k_B \,\sum_{n=1} \bigg\{ a_{n}^{\dagger \: I} a^J_{n } \,\delta_{I J}\,\ln \left(
\sinh^{2}\left(\theta _{n}\right)\right) - a^I_{n} a_{n}^{\dagger \: J
} \,\delta_{I J}\, \ln \left( \cosh^{2}\left(\theta _{n}\right)\right)\bigg\}.
\label{k}
\end{eqnarray}
in the state $\left.\left|0(\theta)\right\rangle\! \right\rangle$ \footnote{This expression comes from the conventional formula for the Von-Neumann entropy in terms of the number operator: $K \equiv -k_B\sum_{n=1} N_{n} \ln N_{n}$ (introduced in Ref. \cite{tu}). The number operator is canonically defined by $
N_{n}= a_{n}^{\dagger \: I}  a_{n }^I$
whose expectation value is proportional to $\sinh^2
\theta_n$.}. In the canonical ensemble, the state (\ref{distr-equil}) is obtained by minimizing the free energy
\begin{equation}
F= \left\langle\! \left\langle 0(\theta) \right|\right. H \left.\left|0(\theta)\right\rangle\! \right\rangle - \frac{1}{\beta } \left\langle\! \left\langle 0(\theta) \right|\right.K \left.\left|0(\theta)\right\rangle\! \right\rangle\label{f}
\end{equation}
with respect to the transformation's parameters $\theta$`s, with $\beta$ constant \cite{tu}.
This state corresponds to the AdS black Hole (\ref{BHstate}) in this free-field description.


It is easy to see that the spacetime states are not
eigenstates of the original Hamiltonian but they are
eigenstates of the combination:
\begin{equation}\label{hamiltonian}
{\widehat H} = H -{\widetilde H},
\end{equation}
in such a way that ${\widehat H}$ plays the r\^{o}le of the
Hamiltonian, generating  temporal translation in the doubled
(\textbf{$CFT^2$}) Fock space. This is then the Hamiltonian which governs the dynamical evolution of the spacetime and its (stringy) excitations.

If we define an effective time independent Hamiltonian of interaction, added to the decoupled CFT$^2$ one, $\widehat{H}$, precisely as being
 a functional of the Bogoliubov generator, we get a good candidate to a model that effectively describes that collapse process: since $G$
  generates the canonical transformations, the main properties of such a model are that the CFT$^2$ algebra of operators, and the states of
   the theory, are preserved; any initial stationary states of $\widehat{H}$ is evolved into another stationary configuration; and the evolution is unitary.
 Actually, the real physical process could violate some of these ideal properties, but our goal here is to argue that an unitary example of dynamics there exists indeed (see remark \textbf{(ii)}).

So naively, we can define the simplest interaction Hamiltonian as
\be\label{Hint}
H_I \equiv \,i \, \epsilon_T(t) \,\sum_n\, \lambda_n\,\, \G_n \,\,,
\ee
where \be \G_n = -i
\,\delta_{I J}\,\sum_{n}\,(a_{n}^{I} \tilde{
a}_{n}^{J}\, - \,\tilde{a}_{n}^{\dagger\:I} a_{n}^{\dagger\:J} )\, ;
\ee
 $\lambda_n$ are effective coupling constants which may be assumed to be small;
 and the step function $\ep_T (t) $ is defined to be $1$ in the interval $(0,T)$ and vanishing otherwise.
 According to this particular dynamics, a growing number of pairs of excitations is created from the vacuum as composites
 $a_{n}^{\dagger\:I} \tilde{a}_{n}^{\dagger\:I}\,\left.\left|0\right\rangle \! \right\rangle $,
  whose component particles locate in each of the two disconnected parts of the spacetime.

The free hamiltonian is $\hat{H}$ that commutes with $H_I$ for construction, then
in the interaction picture, the evolution from the initial state prepared as: $ \left.\left|\psi(t=-\infty)\right\rangle\! \right\rangle \equiv \left.\left|0\right\rangle \! \right\rangle $ and its evolution after a time interval grater than $T$ results:
\be \left.\left|\psi(t>T)\right\rangle\! \right\rangle =
e^{-i  \int_{-\infty}^t \, H_I \,dt} \left.\left|\psi(t=-\infty)\right\rangle\! \right\rangle =
e^{ T\, \sum_n \lambda_n\,\Gamma_n} \left.\left|\psi(t=-\infty)\right\rangle\! \right\rangle \label{evol}. \ee
 By fixing the couplings to be $\lambda_n T \equiv \theta_{n}^{(eq)} $, where $\theta_{n}^{(eq)} $ is given by (\ref{distr-equil}), this formula agrees with (\ref{tva2}), which shows that the final state is the Black Hole (Eq. (\ref{BHstate})).

The trick of doubling CFT to construct the geometries from entanglement, hints to understand unitary gravitational collapse in AdS. This model captures the main ingredients.

Actually, we do not know what precise microscopic/quantum processes are activated in the real Black Hole formation: from a strings perspective, most probably high energy string excitations shall be involved; higher order effective corrections to GR dynamics should be concerned to describe the space time evolution, and complicated initial/boundary conditions should be specified in order to fix uniquely the solution.
However, the example above shows that there exist simple descriptions/models for this process, which may be interpreted as effective dynamics that unitarily evolves a simple initial state of the spacetime (disconnected) into a AdS Black Hole, preserving the CFT structure.

~


Let us \emph{qualitatively} describe the process of formation of an AdS black hole according to this particular dynamics.
This may be more or less organized in the following stages:

~

\begin{figure}
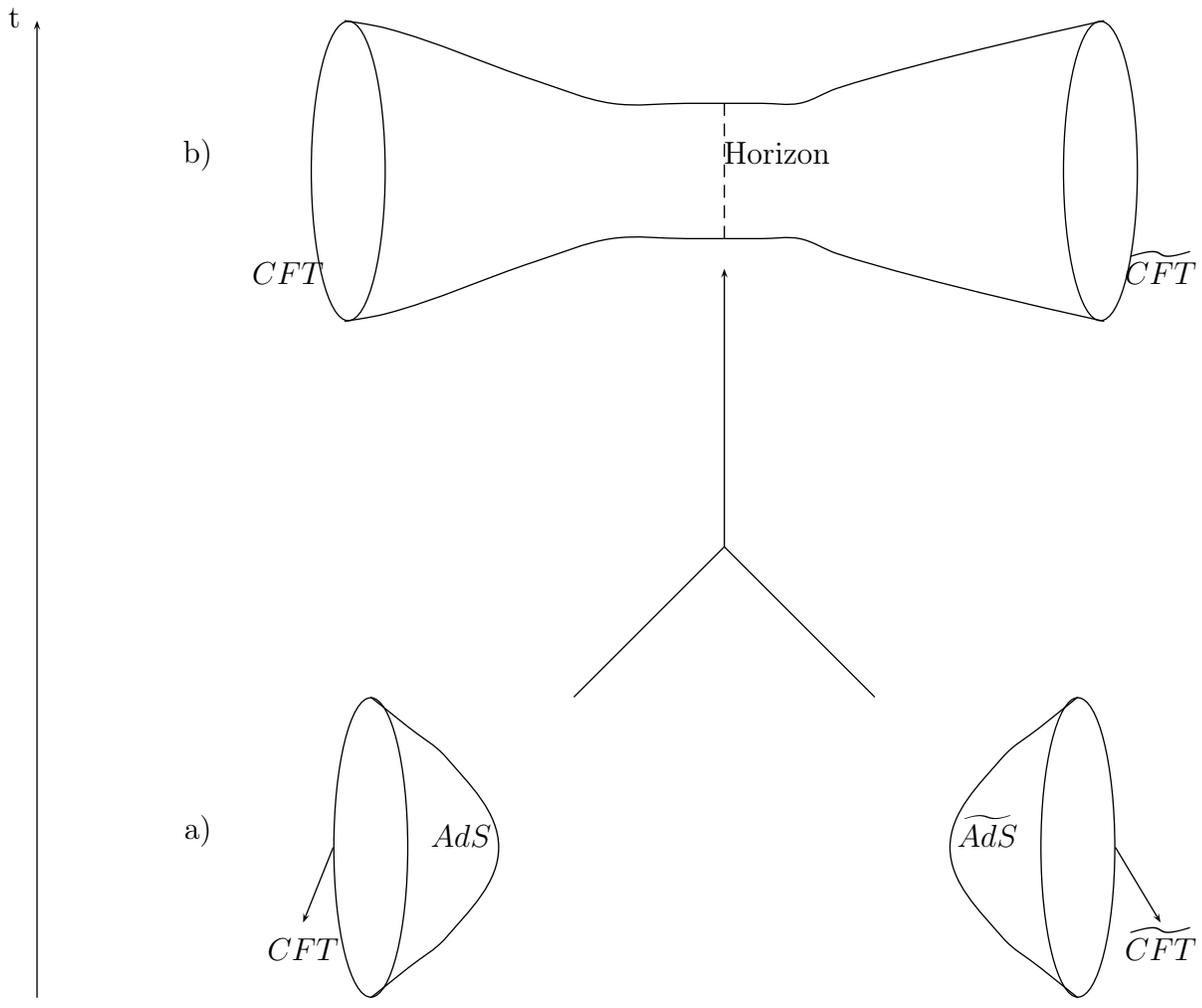

\psline[linewidth=.5pt]{->}(0,-10)(0,3)
\rput[bI](-0.3,2.9){t}
\hspace{3cm}
\psellipse[linewidth=0.5pt](1,1)(0.5,2)
\pscurve[linewidth=.5pt](0.95,3)(1.5,2.9)(3.5,2.2)(4.5,1.9)(5.5,1.9)(6,1.9)(6.5,1.9)(7,1.9)(7.5,2.1)(11.05,3)
\psellipse[linewidth=0.5pt](11,1)(0.5,2)
\pscurve[linewidth=.5pt](0.95,-1)(1.5,-0.9)(3.5,-0.2)(4.5,0.1)(5.5,0.1)(6,0.1)(6.5,0.1)(7,0.1)(7.5,-0.1)(11.05,-1)
\psline[linewidth=.5pt,linestyle=dashed](6,0.1)(6,1.9)
\rput[bI](6.7,1.1){Horizon}
\psline[linewidth=.5pt]{<-}(6,-0.3)(6,-4)
\psline[linewidth=.5pt](6,-4)(4,-6)
\psline[linewidth=.5pt](6,-4)(8,-6)
\psellipse[linewidth=0.5pt](1.3,-8)(0.5,2)
\psellipse[linewidth=0.5pt](10.7,-8)(0.5,2)
\pscurve[linewidth=0.5pt](1.3,-6)(1.8,-6.4)(2.3,-6.8)(3,-8)(2.3,-9.2)(1.8,-9.6)(1.3,-10)
\pscurve[linewidth=0.5pt](10.7,-6)(10.2,-6.4)(9.7,-6.8)(9,-8)(9.7,-9.2)(10.2,-9.6)(10.7,-10)
\psline[linewidth=.5pt]{->}(0.8,-8)(0.4,-9)
\psline[linewidth=.5pt]{->}(11.2,-8)(11.8,-9)
\rput[bI](0.2,-0.5){$CFT$}
\rput[bI](11.8,-0.5){$\widetilde{CFT}$}
\rput[bI](2.5,-8){$AdS$}
\rput[bI](9.5,-8){$\widetilde{AdS}$}
\rput[bI](-1,-8){a)}
\rput[bI](-1,1){b)}
\rput[bI](0.4,-9.5){$CFT$}
\rput[bI](11.8,-9.5){$\widetilde{CFT}$}
\vspace{12cm}
\caption{(a) {\sf Initial state}: The spacetime consists in two disconnected exact AdS geometries, which may contain string excitations.
 It corresponds to a $CFT^2$ state with vanishing entanglement entropy. The figure schematically shows the spacial section of both components of the spacetime
 (b) {\sf Final state}: A Black Hole configuration: the figure shows a spacial slice of the maximally extended Black Hole.}
\label{figure}
\end{figure}

\textbf{S1} The initial state is some field configuration on the a spacetime given by the union
 of two disconnected AdS spacetimes, or "drops" (see Figure \ref{figure}(a)). This is an stationary state, an eigenstate
  of the total hamiltonian $\hat{H}=H - \widetilde{H}$, and the entanglement entropy is vanishing  (more general, pure, "initial'' configurations are constructed in Section 6)

\textbf{S2} In intermediate stages, some effective non-local interaction between the fields of each drop (or, between the sectors $CFT/\widetilde{CFT}$) activates, so as in the toy model (\ref{Hint}). In other words, coupling terms added to $\widehat{H}=H - \widetilde{H} + \lambda H_{int}$ turns out to be important. In the gravity description this must be supported by spacetimes whose asymptotic regions are \emph{causally} connected, as argued in \cite{AdSWH},
which is generically associated to a wormhole geometry connecting the two drops by a throat, as a sort of intermediate configuration.

\textbf{S3} The final (out) state is again (Fig. \ref{figure}(b)) an eigenstate of $\widehat{H}=H - \widetilde{H}$, stationary, given by (\ref{BHstate}), that describes a connected (by non-causal curves) manifold: the maximally extended Black Hole state (Fig. 2).

This process is unitary, reversible, and is schematically very similar to a \emph{tree diagram} as we can see in Fig.\ref{figure}. This suggests a sort scattering matrix description of the quantum process of formation of a Black Hole \cite{smatrix}.
 Let us point out nevertheless, that according to the remark (iii), an entropic principle rules out the process generically, independently from the dynamical details of the specific microscopical model. The model (\ref{Hint}) is only a particularly simple candidate to describe a microscopic dynamics which unitarily evolves the state $\left|0\right\rangle \times \left|\tilde{0}\right\rangle$ into the black hole state (\ref{BHstate}), however the final state of collapse process is independent of the existence or specific form of $H_I$.

\section{Hawking-Page transition}

As an example of application of emergence rule \textbf{I.1}-\textbf{I.2}, the Hawking-Page transition \cite{haw-page} may be described in a very simple way, through a nice quantum mechanical interpretation: as a critical behavior of the quantum amplitude of the state of the spacetime (\ref{geometrystate}). In fact the BH state in the quantum mechanical theory $CFT^2$ is (\ref{BHstate}), at low temperatures, this becomes
\be \left.\left|0(\beta)\right\rangle \! \right\rangle =  \,\frac{e^{-\b E_0}}{Z^{1/2}} \left|0\right\rangle \otimes \left|\tilde{0}\right\rangle + \delta(\beta) \left.\left|\xi(\beta)\right\rangle \! \right\rangle
\label{BHstate-low}, \ee
where $\xi$ is a state orthogonal to $\left|0\right\rangle \otimes \left|\tilde{0}\right\rangle$, and $|\delta(\b)|^2\ll 1$ as $\b \ll E_0^{-1} \sim b$ (here $b$ is the radius of curvature of the anti-de Sitter space). So at low temperatures, the probability of measure and collapse this state into two disconnected AdS spaces (in the fundamental state) is very high, compared with other geometry states.

 Since a more thermodynamical/TFD point of view: the entropy operator $S$ is an observable, and the third law implies that its expectation value vanishes in the limit $T\to 0$; so that, upon observation, the system shall collapse into some disentangled state $\left|\psi\right\rangle \otimes |\,\widetilde{\psi}\,\rangle\,$ in this limit. So the probability of observe and collapse the state (\ref{BHstate-low}) upon proper measurements, into two \emph{disconnected} asymptotically-AdS geometries is very high as $T\to 0$; and in particular, the global AdS pair $\left|0\right\rangle \otimes \left|\tilde{0}\right\rangle$ is highly more probable than other excited states of the AdS geometries.

\section{Bulk fields and perturbative approach}

The main goal of this section is to construct a quantitative approach to this framework in order to show, in particular, how the formal statements \textbf{I}...\textbf{II}, on the structure of the states of the spacetime geometry, looks like in terms of quantum bulk fields in a free-field approximation; and furthermore, to introduce a canonical framework that realizes the model of collapse proposed above.


Let us consider the full Quantum Field Theory defined on a AdS background. This supposedly includes the gravitational modes (deviations from the AdS metric), so it contains Quantum Gravity on this fixed background.  Let us denote as $\phi^J$ all these fields defined on $M_0$; for instance, derived from a level expansion of the Type II \textbf{B} String Theory (the bosonic sector)\footnote{The index $J$ synthetically denotes all the species and tensor types of the bulk fields}. AdS/CFT states that the (bulk) Hilbert space of fields, quantized on a global AdS background geometry, ${\cal H}_{bulk}$, and the CFT Hilbert space ${\cal H}_{CFT}$ are equivalent, this
implies that the set of configurations $\phi^J$ on a effective Cauchy surface \cite{Avis} of $M_0$ (which constitutes a basis of ${\cal H}_{bulk}$) is in a \emph{one-to-one} correspondence with a complete basis $\{\left|n\right\rangle\}_n$ of the CFT Hilbert space.

In principle, the construction we are going to set here is similar for all the bulks fields $\phi^J$, but we may have a good (quantitative) description of this process just for looking the feature of a real scalar mode. So for technical simplicity, let us consider a free scalar field $\phi$ of mass $m$ on the global $AdS_{5} \times S^5$ background, considered fix; we assume, in addition, that this field is free and effectively decoupled from other bulk fields asymptotically, so as in a conventional scattering (in effective Cauchy surfaces so as $t\to \pm \infty$, and near the conformal boundaries) \cite{giddings-gary}, and just shall look the sector of the Hilbert space associated to it. We are considering asymptotic observers which realize measures and collapse states belonging to this space.

Then, the free scalar field may be canonically quantized \cite{Avis}. The Klein-Gordon equation solutions express as ${\sf \phi}(t,{\Omega},x)=e^{-i\omega
t}Y_{lm}({\Omega})f_{l\omega}(x) $, where the AdS$_{d+1}$ manifold is fully covered by the so-called global coordinates
where the metric takes the form\footnote{We use the coordinate $x=\tanh\rho$, where $\rho$ is the standard radial variable, in order
to map the conformal boundary to $x=1$.}
\be\label{adsglobal}
ds^2=R^2\left[-\frac{dt^2}{1-x^2}+\frac{dx^2}{(1-x^2)^2}+\frac{x^2}{1-x^2}d\Omega^2_{d-1}\right].
\ee
We quantize this field by considering just the normalizable modes:
\be
 \omega_{kl}=\pm(2k+l+\Delta_+)~,~~~ \Delta_+=\frac{d}{2}+\mu~~,~~~\mu=\sqrt{\frac{d^2}{4}+m^2R^2}\,,~~~k,l=0,1,2\ldots\,,
\label{modos}
\ee
which is interpreted as dual to the CFT stationary states defined on the $S^3\times \mathbb R$ conformal boundary
of AdS. The discreteness of this spectrum manifests the ``box'' character
of AdS and from the dual perspective arises from the compactness of $S^3$, which is crucial for our argument of unitarity of the Bogoliubov's map (see \textbf{(i), (ii)}, Sec 4).

So the expression for the field operator is:
\be\label{sol}
{\sf \phi}(t,{\Omega},x)= \sum_{klm} \, \left(e^{-i\omega_{kl}
t}Y_{lm}({\Omega})f_{lk}(x) a^\dag_{klm} \right) + h.c.\,, ~~~~~w_{kl}>0\,,
\ee
where the coefficients $a_{klm}, a^\dag_{klm}$ are defined as anihilation/creation operators that satisfy the commutation relations
$[a_n, a^\dag_{n'}] = \delta_{n n'}$ (other commutators vanish), where $n$ denotes the complete set of numbers $k,l,m$.
 Therefore, the bulk Hamiltonian may be canonically expressed in terms of these modes by eq. (\ref{hamcan}).

  Following \textbf{I.1, I.2}, in the spacetime/bulk picture (subsection 3.1), let us take another (disconnected) copy of this system denoted by tilde, namely another global AdS spacetime $\widetilde{M}_0$, described by a metric as (\ref{adsglobal}), and a field operator built according to the TFD rules (\ref{til}):
 \be\label{sol-tilde}
\widetilde{{\sf \phi}}(\tilde{t},\widetilde{{\Omega}},\widetilde{x})= \sum_{klm} \, \left( \, e^{i\omega_{kl} \tilde{t}
}Y^*_{lm}(\widetilde{{\Omega}})f_{lk}(\widetilde{x}) \widetilde{a}^\dag_{klm} \right) + h.c.\,, ~~~~~w_{kl}>0\,,
\ee
The time parameters $t, \tilde{t}$ run independently but they may be chosen to be the same \cite{eternal},
such that the Hamiltonian $\hat{H}=H-\tilde{H}$ generates the time evolution of the total system.
The Fock space is ${\cal F}^2(M_0)\equiv {\cal F} \otimes \widetilde{{\cal F}}$, generated by applying the creation operators $a^\dag_{klm},\widetilde{a}^\dag_{klm}$ to the vacuum state. The standard vacuum in this extended theory is defined by
\be {a}^{J}_{n}\left.\left|0\right\rangle \! \right\rangle= \A_{n}^J
\left.\left|0\right\rangle \! \right\rangle = 0
  , \ee for $w_n>0$ and $\left.\left|0\right\rangle \!
\right\rangle =|0\rangle \otimes|\widetilde{0} \rangle$ as usual.

One may now to consider a family of spacetimes $\left.\left|M(\theta)\right\rangle\! \right\rangle$ (labeled by a set of parameters $\theta$) described in Assumption \textbf{I}, expressed by the formulas (\ref{geometrystate})/(\ref{gstate-G}), such that:

~

(a) $\left.\left|M\right\rangle\! \right\rangle$ admits a perturbative quantization of the $\phi$ mode and the construction of a Fock space ${\cal F}^2(M)(\cong {\cal F} \otimes \widetilde{{\cal F}})$ in the sense explained above;

(b) This state is related to the ground state (a disconnected pair of global AdS spacetimes)
 as:
\be\label{geometrystate-G} \left.\left|M\right\rangle\! \right\rangle =
e^{-i{G}} \left.\left|0\right\rangle \! \right\rangle  ; \ee

(c) $e^{-i{G}}: {\cal F}^2(M_0) \rightarrow {\cal F}^2(M)$ is
 an unitary and canonical map (a Bogoliubov transformation). Namely, it commutes with the hamiltonian $\widehat{H}$ and preserves the commutators algebra.

~

Let us explicitly analyze the structure of these spacetimes in the spirit of the statements \textbf{I.1,2}, described here as ground states of bulk fields. In particular, an AdS Black Hole may be described as a vacuum state of this type, given by equations (\ref{tva2}), and (\ref{distr-equil}) (see Ref. \cite{ads-haw}).

The last requirement (c) reflects the fact that the perturbative description of the
degrees of freedom, so as the quantum transition amplitudes (unitarity), are preserved by $e^{-i{G}}$. The general form of the Bogoliubov transformation that fix
the form of the generator is given by the following relation \cite{chu-ume}
\begin{equation}
\left(
\begin{array}{c}
a^{\prime } \\
\tilde{a}^{\dagger \prime }
\end{array}
\right) =e^{-i\G}\left(
\begin{array}{c}
a \\
\tilde{a}^{\dagger }
\end{array}
\right) e^{i\G}={\cal B}\left(
\begin{array}{c}
a \\
\tilde{a}^{\dagger }
\end{array}
\right) ,\quad \left(
\begin{array}{cc}
a^{\dagger ^{\prime }} & -\tilde{a}^{\prime }
\end{array}
\right) =\left(
\begin{array}{cc}
a^{\dagger } & -\tilde{a}
\end{array}
\right) {\cal B}^{-1},
\label{Bogoliubov}
\end{equation}
where the generator of the transformation, $\G$, called
{\em the Bogoliubov operator} is Hermitian, and then $\cal B$ is a $2 \times 2$ complex matrix
\begin{equation}
{\cal B}=\left(
\begin{array}{cc}
u & v \\
v^{*} & u^{*}
\end{array}
\right) ,   ~~~~~~~~ \qquad \left| u\right| ^{2}-\left| v\right| ^{2}=1,
\label{Bmatrix}.
\end{equation}
 The operators that satisfy these relations have the
following form \cite{chu-ume}
\begin{eqnarray}\label{gen-generators}
\G_{1_{n}}^{} &=&\theta _{1_{n}}\left( a_{n}\cdot \tilde{a}_{n}+\tilde{
a}_{n}^{\dagger }\cdot a_{n}^{\dagger }\right) ,  \nonumber \\
\G_{2_{n}}^{} &=&i\theta _{2_{n}}\left( a_{n}\cdot \tilde{a}_{n}-
\tilde{a}_{n}^{\dagger }\cdot a_{n}^{\dagger }\right) , \nonumber \\
\G_{3_{n}}^{} &=&\theta _{3_{n}}\left( a_{n}^{\dagger }\cdot a_{n}+
\tilde{a}_{n}^{\dagger }\cdot \tilde{a}_{n}+ 1 \right)= \, \theta _{3_{n}}\, (N_n + \tilde{N}_n + 1)~,
\label{generators}
\end{eqnarray}
where $\theta$'s are the real parameters
which, for convenience, have been included in the operators; the dot represents sum over all the physical fields,
 formally labeled by the indices $I, J$. $N_n= a_{n}\cdot a_{n}^{\dagger }$ is the canonical occupation number of the mode $n$. It is
easy to verify that the generators (\ref{generators}) satisfy the
$SU\left( 1,1 \right)$ algebra \be \left[
\G_{i_{n}},\G_{j_{n}}\right] =-i\Theta _{ijk}\G_{k_{n}}\,\,\, ,
\label{su11} \ee where $\Theta _{ijk}\equiv 2\frac{\theta _{i_{n}}\theta _{j_{n}}}{\theta _{k_{n}}}$.

The generators $\G_{3_{n}}^{} $ in the last line,
are associated to creation of particles/excitations in each separated drop.
 It is easy to see that these do not generate entanglement (do not contribute to the entropy).
  The family of states generated by this operator are the pure states of the spacetime
 containing string excitations, it is the set of initial configurations that potentially may collapse into connected spacetimes, with causally hidden regions (as extended black holes).

As we can see from  (\ref{generators}), the most general
symmetry generator $\G=\sum_{n}\left( \G\right)_{n}$ takes the
following form
\be
\G_{n} =\lambda _{1_{n}}\tilde{a}_{n}^{\dagger }\cdot
a_{n}^{\dagger }-\lambda _{2_{n}}a_{n}\cdot \tilde{a}_{n}+\lambda
_{3_{n}} \left( a_{n}^{\dagger }\cdot a_{n}+\tilde{a}_{n}^{\dagger }\cdot
\tilde{a}_{n} + 1\right)
\label{rlgen}
\ee
and the coefficients represent complex linear combinations of $\theta$'s
\begin{equation}
\lambda _{1_{n}}=\theta _{1_{n}}-i\theta _{2_{n}},\quad \lambda
_{2_{n}}=-\lambda _{1_{n}}^{*},\quad \lambda _{3_{n}}=\theta _{3_{n}}.
\label{lambdas}
\end{equation}
These are called generalized Bogoliubov transformations or simply G-transformations \cite{ume,chu-ume} which form a $SU(1,1)$ algebra. Of course, This shall be dual to a symmetry of the composite theory $CFT^2$ we are considering here.


By applying the
disentanglement theorem for $su(1,1)$ \cite{disentanglement}, one can
obtain \emph{the most general} state of the spacetime (\ref{geometrystate}):
\bea
\left.\left|M(\Omega)\right\rangle \! \right\rangle
=\prod_{n}e^{\Omega_{1_{n}}\left( \tilde{a}_{n}^{\dagger }\cdot
a_{n}^{\dagger }\right) }e^{\log \left( \Omega_{3_{n}}\right) \left(
a_{n}^{\dagger }\cdot a_{n}+ \tilde{a}_{n}^{\dagger }\cdot \tilde{a}
_{n}+ f\,\delta_{nn}\right) }e^{\Omega_{2_{n}}
\left( a_{n}\cdot \tilde{a}_{n}\right)}
 \left.\left|0\right\rangle \! \right\rangle\;\;, \label{vacalpha}
\eea
as ground state for the bulk fields $\phi^I$ (annihilated by $a^I_{n}(\theta), \tilde{a}^I_{n}(\theta)$).
The new parameters are given by the relations:
\be
\Omega_{1_{n}}=\frac{-\lambda _{1_{n}}\sinh \left( i\Lambda _{n}\right) }{%
\Lambda _{n}\cosh \left( i\Lambda _{n}\right) +\lambda _{3_{n}}\sinh \left(
i\Lambda _{n}\right) },\quad \Omega_{2_{n}}=\frac{\lambda _{2_{n}}\sinh
\left( i\Lambda _{n}\right) }{\Lambda _{n}\cosh \left( i\Lambda _{n}\right)
+\lambda _{3_{n}}\sinh \left( i\Lambda _{n}\right) },
\label{lambda12}
\ee
\begin{equation}
\Omega_{3_{n}}=\frac{\Lambda _{n}}{\Lambda _{n}\cosh \left( i\Lambda
_{n}\right) +\lambda _{3_{n}}\sinh \left( i\Lambda _{n}\right) }~,\quad\Lambda _{n}^{2}\equiv \left( \lambda _{3_{n}}^{2}+\lambda _{1_{n}}\lambda
_{2_{n}}\right) .
\label{lambda3}
\end{equation}
Since the pure vacuum is annihilated by $a_{n}^{\mu }$
and $\tilde{a}_{n}^{\mu }$, the expression for spacetime states reduces to
\begin{equation}
\left.\left|M(\Omega)\right\rangle \! \right\rangle
=\prod_{n}(\Omega_{3_{n}})^{ f \,\delta_{nn}}
e^{\Omega_{1_{n}}\left( \tilde{a}_{n}^{\dagger
}\cdot a_{n}^{\dagger }\right) }  \left.\left|0\right\rangle \! \right\rangle\,\,\,\,.
\label{thermvacfin}
\end{equation}
The operators defined on global AdS are mapped to entangled ones (according to \textbf{I}, on connected asymptotically AdS spacetimes with two boundaries) by the
corresponding Bogoliubov generators \be a_{n}^{\mu }\left( \theta
\right)  = e^{-i\G_{n}}a_{n}^{\mu }e^{i\G_{n}},\qquad \tilde{a}_{n}^{\mu
}\left( \theta \right) =e^{-i\G_{n}}\tilde{a}_{n}^{\mu }e^{i\G_{n}}.
\label{thermop} \ee Similar relations hold for the creation
operators. The entangled operators satisfy the same canonical
commutation relations as the
 pure operators at $\theta=0$ by virtue of the requirement (c).
Alternatively, one can organize the operators in doublets \cite{ume,chu-ume} and
 represent the Bogoliubov transformation
as
\begin{equation}
\left(
\begin{array}{c}
a_{n}^{\mu }\left( \theta \right)  \\
\tilde{a}_{n}^{\mu \dagger }\left( \theta \right)
\end{array}
\right) ={\cal B}_{n}\left(
\begin{array}{c}
a_{n}^{\mu } \\
\tilde{a}_{n}^{\mu \dagger }
\end{array}
\right) ,
\label{doublet}
\end{equation}
where the explicit form of the ${\cal B}_n$ matrices is given by
\begin{equation}
{\cal B}_{n}=\cosh \left( i\Lambda _{n}\right) {\Bbb I} +\frac{\sinh \left(
i\Lambda _{n}\right) }{\left( i\Lambda _{n}\right) }\left(
\begin{array}{cc}
i\lambda _{3_{n}} & i\lambda _{1_{n}} \\
i\lambda _{2_{n}} & -i\lambda _{3_{n}},
\end{array}
\right)
\label{explB}
\end{equation}
where $\Bbb I$ is the $2 \times 2$ identity matrix.


~

\textbf{Remark:} this simplified framework provides a manifest description of what the states of the spacetime \emph{are} in the context of the proposal \textbf{I}. In fact, the state elements of the Fock space annihilated by all the operators $a_n^J(\theta), \tilde{a}_n^J(\theta)$ mathematically defines $\left.\left|M(\theta)\right\rangle\! \right\rangle$.

~

Following the assumptions \textbf{II}, the entropy operator is defined such that its average value be proportional to
the
entropy of the bosonic field at thermal equilibrium divided by the Boltzmann's
constant
\cite{tu}. The entropy of the bosonic field can be computed
as the
expectation value of the entropy operator in the entangled vacuum
\begin{equation}
\frac{1}{k_{B}}\left\langle \!\left\langle 0\left( \theta \right) \right|
\right. K\left. \left| 0\left( \theta \right) \right\rangle \!\right\rangle
=\left\{ \sum_{n}\left[ \left( 1+N_{n}\right) \log \left( 1+N_{n}\right)
-N_{n}\log \left( N_{n}\right) \right] \right\},
\label{entrboson}
\end{equation}
where $N_n= a_{n}\cdot a_{n}^{\dagger }$ is the canonical occupation number of the mode $n$. This manifestly measures the \emph{entanglement} entropy between local degrees of freedom placed in the two causally disconnected regions of the space $M(\theta)$ (it shall be investigated in connection with the holographic proposal \cite{takaya}). In this context, the entropy operator may be expressed as:
\be
K=-\sum_{n}\left[ a_{n}^{\dagger }\cdot a_{n}\log \left( \frac{
\lambda _{1_{n}}\lambda _{2_{n}}}{\Lambda _{n}^{2}}\sinh ^{2}\left( i\Lambda
_{n}\right) \right) -a_{n}\cdot a_{n}^{\dagger }\log \left( 1+\frac{\lambda
_{1_{n}}\lambda _{2_{n}}}{\Lambda _{n}^{2}}\sinh ^{2}\left( i\Lambda
_{n}\right) \right) \right] \label{rightentr}
\ee
where
\begin{equation}
N_{n}= \left[ \frac{\lambda _{1_{n}}\lambda _{2_{n}}}{\Lambda _{n}^{2}}
\sinh^{2}\left( i\Lambda _{n}\right) \right] ,\,\,\,\,\,\left\langle \!\left\langle 0\right| \right. \tilde{a}_{n}\cdot \tilde{a}
_{n}^{\dagger }\left. \left|0\right\rangle \!\right\rangle \equiv 1 .
\label{nnumb}
\end{equation}
One obtains the thermodynamic entropy as the average of this operator in the vacuum, identified with the state of the spacetime:
\be
S =k_{B}\left\langle \!\left\langle M\left( \theta \right) \right| \right.
K\left. \left| M\left( \theta \right) \right\rangle \!\right\rangle.
\label{closedstringentr}
\ee
This quantity may be explicitly maximized with respect to the parameters $\theta$ in order to obtain the family of stationary spacetimes defined by (a), (b), and (c).

\section{Concluding remarks}

In this initial work we have shown that
in quite general conditions, namely, in quantum gravitational theories on a background geometry dual to certain field theories on finite volume, a Black Hole state (Eq. (\ref{BHstate})) with the same asymptotics, may be achieved from a pure state as $|0\rangle \otimes|\widetilde{0} \rangle$ (or disentangled excitations\footnote{Those vacua generated by $\G_{3_{n}}^{} $ (eq. (\ref{su11})) in the free-field picture.}) by an unitary transformation.
This is a Bogoliubov transformation, which may be seen as the simplest example of an evolution operator that realizes the collapse process.
Non surprisingly, the particular structure of the AdS space (via the Maldacena conjecture) played a crucial role to argue unitarity. The TFD/duplication point of view is the new ingredient to understand some dynamics (and states) of the spacetime, and reveals promising aspects towards a S-matrix description.

In fact, this scenario shed light on the elusive emergence mechanism/interpretation: one get the space of states of the spacetime by considering the quantum statistical information of the CFT theory in the extended Hilbert space $\sim {\cal H}_{CFT}^2$, that encodes the thermodynamics of the system in agreement with remarkable observations \cite{macrogeom}. In this sense, one may interpret the gravitational collapse as a current thermodynamical process ruled out by conventional thermodynamic laws, suggesting an universal principle (as observed in \textbf{(iii)}, Sec. 4 and formulated in \textbf{II.1} $\&$ \textbf{II.2}).
The next step is to recover the semiclassical notion of spacetime from states (\ref{gstate-G}), and to derive the GKPW holographic map (Ref. \cite{GKPW}) in the context proposed here \cite{forth}.

\section{Acknowledgements}

   L. Alvarez-Gaume is specially acknowledged for discussions, and stimulating comments from the beginning.
    The author is also grateful to J. Russo for interesting observations, R. Arias, J. A. Helayel-Neto, and A. Santana
  for useful correspondence. This work was partially supported by: CONICET PIP 2010-0396 and ANPCyT PICT 2007-0849.

\end{document}